\documentclass[aps,twocolumn,showpacs]{revtex4}
\usepackage{dcolumn}
\usepackage{graphicx}
\usepackage{amsmath}
\usepackage{amsfonts}
\usepackage{amssymb}
\usepackage{psfrag}
\usepackage{wrapfig}
\usepackage{subfigure}
\usepackage{makeidx}
\usepackage{bm}
\usepackage{epsf}
\newtheorem{thm}{Theorem}
\newtheorem{lem}{Lemma}

\begin{document}

\title{High-order Rogue Wave solutions for the Coupled Nonlinear Schr\"{o}dinger Equations-II}
\author{Li-Chen Zhao$^{1}$, Boling Guo$^{2}$, }
\author{Liming Ling$^{3}$ }\email{Corresponding-author(linglm@scut.edu.cn)}
\address{$^1$Department of Physics, Northwest University, Xi'an 710069, China}
\address{$^2$Institute of Applied Physics and Computational Mathematics,
Beijing 100088, China}
\address{$^3$School of Mathematics, South China University of Technology,
Guangzhou 510640, China}

\date{November 6, 2014}
\begin{abstract}
We study on dynamics of high-order rogue wave in two-component coupled nonlinear Schr\"{o}dinger equations. Based on the generalized Darboux transformation and formal series method, we obtain the high-order rogue wave solution without the special limitation on the wave vectors.  As an application, we exhibit the first, second-order rogue wave solution and the superposition of them by computer plotting. We find the distribution patterns for vector rogue waves are much more abundant than the ones for scalar rogue waves, and also different from the ones obtained with the constrain conditions on background fields. The results further enrich and deep our realization on rogue wave excitation dynamics in such diverse fields as Bose-Einstein
condensates, nonlinear fibers, and superfluids.

\end{abstract}

 \maketitle

\section{Introduction}
Vector rogue wave(RW) has been paid much attention, since the components in nonlinear systems are usually more than two and the dynamics of vector RW demonstrate many striking dynamics in contrast to the ones in scalar system. Firstly, there are some new excitation patterns for vector RW, in contrast to the well-known eye-shaped one in scalar system \cite{Ruban,N.Akhmediev,Kharif,Pelinovsky}.  For example,  dark RW was presented numerically \cite{Bludov} and analytically
\cite{zhao2,Chen,Baronio1} for two-component coupled systems. Four-petaled RW was reported recently in three-component
coupled systems \cite{Zhao3,Degasperis}, and even two-component coupled system \cite{Zhao4}. Secondly, the number of RW in temporal-spatial distribution plane is also different from the ones in scalar systems. We have demonstrated that two or four fundamental RWs can emerge on the distribution plane \cite{zhao2,lingzhao}, which is absent for scalar systems \cite{Yang,Ling,He,Ling2,Akhmediev}.

It should be noted that the two or four RW patterns were derived under some certain constrains on background fields \cite{zhao2,lingzhao}. The relative wave vector for the two background fields should satisfy a certain relation with the background amplitudes. Then, are the constrain  conditions essential for these new distribution patterns? We revisit on high-order RWs by developing the deriving method. Furthermore, the RW patterns for four RW pattern or six RW pattern are all eye-shaped one for the coupled model with constrain conditions. Considering the four-petaled RW and anti-eye-shaped RW can exist in the system, one could expect that these different pattern can coexist for some special superposition cases. Therefore, it is essential to find out how they can be superposed and the properties of the superposition.

In this paper,  we present a method to derive high-order RW solution with releasing the constrains conditions on background fields. We find that two or
four fundamental RWs can emerge for the superposition of first-order RW with itself or the second-order vector RW in the
coupled system without the constrain conditions before. Especially, one four-petaled RW can coexist with three eye-shaped ones and other different pattern combination can exist in the coupled system, in contrast to the four eye-shaped ones coexisting case reported before.
Moreover, six RWs can emerge on the distribution
for the superposition of two second-order vector RWs, which can be constituted of different fundamental RW patterns. For example, three four-petaled and three eye-shaped ones can coexist on the distribution plane. These results are different from the ones in our previously reported results \cite{lingzhao}, and could further enrich our knowledge on RW excitation dynamics in many different coupled systems.

\section{Generalized Darboux transformation and rogue wave formula}
In this section, we firstly recall some basic knowledge of generalized
Darboux transformation. We consider the following focusing coupled
Schr\"oinger equation(NLSE)
\begin{equation}\label{cnls}
    \begin{split}
      i q_{1,t}+\frac{1}{2}q_{1,xx}+(|q_1|^2+|q_2|^2)q_1 &=0,  \\
      i q_{2,t}+\frac{1}{2}q_{2,xx}+(|q_1|^2+|q_2|^2)q_2 &=0.
    \end{split}
\end{equation}
The coupled NLSE can be used to describe evolution of localized waves in a two-mode nonlinear fiber, two-component Bose-Einstein condensate, and other coupled nonlinear systems \cite{Baronio,Guo}. It admits the following Lax pair:
\begin{equation}\label{Lax}
    \begin{split}
       \Phi_x &=U(\lambda;Q)\Phi, \\
       \Phi_t &=V(\lambda;Q)\Phi,\\
    \end{split}
\end{equation}
where
\begin{equation*}
\begin{split}
    U(\lambda;Q)&=i\lambda (\sigma_3+I_3)+Q,  \\
    V(\lambda;Q)&=i\lambda^2(\sigma_3+I_3)+\lambda Q+\frac{i}{2}\sigma_3(Q^2-Q_x)+ic\ I_3 \\
   Q &=\begin{pmatrix}
        0 & i\bar{q}_1 & i\bar{q}_2 \\
        iq_1 & 0 & 0 \\
        iq_2 & 0 & 0 \\
      \end{pmatrix}
    ,\,\left.
         \begin{array}{ll}
           \sigma_3&=\mathrm{diag}(1,-1,-1),
         \end{array}
       \right.
\end{split}
\end{equation*}
$I_3$ is a $3\times 3$ identity matrix and the symbol overbar represents complex conjugation. The parameter $c$ is a real constant.
The compatibility condition $\Phi_{xt}=\Phi_{tx}$ gives the CNLS \eqref{cnls}.

We can convert the system \eqref{Lax} into a new system
\begin{equation}\label{cnls}
    \begin{split}
   \Phi[1]_x &=U(\lambda;Q[1])\Phi[1],\\
       \Phi[1]_t &=V(\lambda;Q[1])\Phi[1],
        \end{split}
\end{equation}
 by the following elementary
Darboux transformation
\begin{equation}\label{DT}\begin{split}
                             \Phi[1]&=T\Phi,\,T=I+\frac{\bar{\lambda}_1-\lambda_1}{\lambda-\bar{\lambda}_1}\frac{\Phi_1\Phi_1^{\dag}}{\Phi_1^{\dag}\Phi_1}, \\
                              Q[1]&=Q+i(\bar{\lambda}_1-\lambda_1)[P_1,\sigma_3],
                          \end{split}
\end{equation}
where $\Phi_1$ is a special solution for system \eqref{cnls} at $\lambda=\lambda_1$.
By the way, if the solution $\Phi_1$ is a nonzero solution, the transformation is nonsingular about $(x,t)\in \mathbb{R}^2$.
If there exists $\Phi_1^{\dag}\Phi_1|_{x=x_0,t=t_0}=0$, then $\Phi_1(x_0,t_0)=0$. By the existence and uniqueness theorem of ODE, we can obtain
$\Phi(x,t)=0$. This is a contradiction. Thus any nonzero solutions could keep the non-singularity of DT. It follows that the solutions of CNLS are non-singularity.

To obtain the general
RW solutions for CNLS, we choose the general plane wave solution
\begin{equation}\label{seed}
    q_1[0]=a_1 \exp{\theta_1},\,\,q_2[0]=a_2 \exp{\theta_2},
\end{equation}
where $\theta_1=i\left [b_1x+(a_1^2+a_2^2-\frac{1}{2}b_1^2)t\right]$, $\theta_2=i\left [b_2x+(a_1^2+a_2^2-\frac{1}{2}b_2^2)t\right]$. Since CNLS equation possesses Galileo symmetry, then we can set $b_2=-b_1.$ It has been shown that the relative wave vector $2 b_1$ can induce the RW pattern transition \cite{Zhao4}.
We firstly investigate the fundamental solution of Lax pair with the general plane wave solution to develop a new method for deriving new RW solutions. Substituting seed solution \eqref{seed} into equation
\eqref{Lax}, we can obtain a vector solution of the Lax pair with $c=a_1^2+a_2^2$
\begin{equation*}
    \Phi(\lambda)=D\begin{bmatrix}
             \exp{\omega}
             \\[4pt]
             \displaystyle{\frac{a_1\exp{\omega}}{\chi+b_1}} \\[8pt]
             \displaystyle{\frac{a_2\exp{\omega}}{\chi-b_1}} \\
           \end{bmatrix}
\end{equation*}
where
\begin{equation*}
    \begin{split}
      D&=\mathrm{diag}\left(1,\,{\rm e}^{\theta_1},\,{\rm e}^{\theta_2}\right) \\
      \omega&=i\left[\chi x+\frac{1}{2}\chi^2t\right]
    \end{split}
\end{equation*}
and $\chi$ is a multiple root for the following cubic equation
\begin{equation}\label{cubic}
    {\xi}^{3}-2\,\lambda\,{\xi}^{2}-\left(a_1^{2}+a_2^{2}
+b_1^{2} \right) \xi+(a_1^{2}-a_2^{2})b_1+2b_1^{2}\lambda=0.
\end{equation}
If the above cubic equation \eqref{cubic} possesses a multiple root, there exist rogue wave solutions. In other words, the spectral parameter
$\lambda_i$ must be satisfied the following quartic equation (discriminant equation for equation \eqref{cubic}):
\begin{equation}\label{discrim}
\begin{split}
   &b_1^{2}\lambda^{4}+ \frac{1}{2}\left(a_1^{2}-
a_2^{2} \right)b_1 \lambda^{3}\\&- \left[\frac{1}{2}b_1^{4}- \frac{5}{4}\left(a_1^{2}+a_2^{2}
 \right)b_1^{2}-\frac{1}{16}\left(a_1^{2}+a_2^{2}
 \right)^{2}
\right] \lambda^{2}\\&+{\frac {9b_1\left(a_1^{2}+a_2^{2}-2b_1
^{2} \right)\left( a_1^{2}-a_2^{2} \right) }{16}}
\lambda\\&+\frac{1}{16}b_1^{6}+\frac{3}{16}\left(a_1^{2}+a_2^{2}
 \right) b_1^{4}+\frac{1}{16}\left(a_1^{2}+a_2^{2}\right) ^{3
}\\&-{\frac {3\left( 5a_1^{2}-a_2^{2} \right)\left( a_1^{2}-5a_2^{2}\right)
b_1^{2}}{64}}=0.
\end{split}
\end{equation}
It is readily to see that the discriminant for quartic equation \eqref{discrim} is
\begin{equation*}
    \Delta=\frac{a_1^2a_2^2b_1^2}{16384}\left[(a_1^2+a_2^2+4b_1^2)^3-27a_1^2a_2^2(4b_1^2)\right]\geq 0.
\end{equation*}
When $\Delta=0$, this is the degenerate case, which have been researched in the previous work \cite{lingzhao}. We found that four or six RWs can exist in the coupled system with some additional constrains on relative wave vector and background amplitudes.  In this work, we consider the non-dengenerate case $\Delta>0$ which can relax these constrain conditions, and try to find if there are some new patterns for RW excitations in the coupled system.
In this case, the discriminant \eqref{discrim} possesses two pairs of conjugation complex roots for the fixed parameters $a_1$, $a_2$ and $b_1$ (In general, when $\Delta>0$, the quantic equation \eqref{discrim} could possesses four real roots. But in this case, we can prove that the quantic equation \eqref{discrim} merely has two pairs of conjugation complex roots with a similar way given in \cite{ling-zhao-guo}). We denote two pairs of
conjugation complex roots as $\lambda_i$ and $\bar{ \lambda_i}$($i=1,2$). The corresponding double roots for equation \eqref{cubic} are $\chi_i$ and $\bar{\chi_i}$ respectively. Since the dynamics of RW solutions with conjugate roots are similar just with RW location difference, we just consider $\lambda_i$ and $\chi_i$ cases.

In this paper, we use the formal series to tackle with high-order rational solutions for CNLS. Since the Kadan formula is complicated to derive for the matrix whose eigenvalue equation is a high-order one, and it is not easy to derive high-order RW solution from Kadan form solutions, we would like to use the asymptotical series to replace the Kadan formula.
We have the following lemmas:
\begin{lem}\label{lem1}
The formal series
\begin{equation}\label{asymp}
\begin{split}
  \lambda_i(\epsilon_i)&=\lambda_i+{d_i}\epsilon_i^{2},\,\,  d_i=\frac{2\lambda_i-3\chi_i}{2b_1^2-2\chi_i^2},  \\
  \chi_i(\epsilon_i)&=\sum_{j=0}^{\infty}\chi_i^{[j]}\epsilon_i^j,\,\, \chi_i^{[0]}=\chi_i,
\end{split}
\end{equation}
satisfy the cubic equation \eqref{cubic}.
The $\epsilon_i$ is small complex parameter,
The parameters $\chi_i^{[1]}=1$ and $\chi_i^{[j\geq 2]}$ can be determined recursively
\begin{equation*}
\begin{split}
&\chi_i^{[j-1]}\\
 =&\frac{1}{(3\chi_i-\lambda_i)}\left[\sum_{\substack{m+n+k=j\\
  0\leq m,n,k\leq j-2}}\frac{-\chi_i^{[m]}\chi_i^{[n]}\chi_i^{[k]}}{2}\right. \\
   &\left.+\sum_{\substack{m+n=j\\
  0\leq m,n\leq j-2}}\lambda_i\chi_i^{[m]}\chi_i^{[n]}+\sum_{\substack{m+n=j-2\\
  0\leq m,n}}d_i\chi_i^{[m]}\chi_i^{[n]}\right],\,\, j\geq3.
\end{split}
\end{equation*}
\end{lem}
Specially, we have
\begin{equation*}
    \begin{split}
      \chi_i^{[2]}=&{
\frac{\left( {b_1}^{2}-4\lambda_i\chi_i+5\chi_i^{2} \right)
}{ 2\left( 2\,\lambda_i-3\,\chi_i \right)  \left( {{  b_1}}
^{2}-{\chi_i}^{2} \right) }},\\
     \chi_i^{[3]}=&{\frac{\substack{2\,\chi_{{i}}^{4}+4\, \left( a_{{1}}^{2}+a_{{2}}^{2}+3\,b_{{1}}
^{2} \right)\chi_{{i}}^{2}+8\,b_{{1}} \left( a_{{1}}^{2}-a_{{2}
}^{2} \right) \lambda_{{1}}+3(a_{{1}}^2+a_{2}^2)^{2}+2\,b_{{4}}^{4}
}}{ 8\left(b_1^2-\chi_i^2
 \right)^{2}\left(2\lambda_i-3
\chi_i\right) ^{2}}}.
    \end{split}
\end{equation*}

It follows that
\begin{equation*}
   \omega_i=i\left[\chi_i x+\frac{1}{2}\chi_i^2t\right]=\sum_{k=0}^{\infty}X_i^{[k]}\epsilon_i^{k}
\end{equation*}
where
\begin{equation*}
  X_i^{[k]}=i\left(\chi_i^{[k]}x+\frac{1}{2}\sum_{j=0}^{k}\chi_i^{[j]}\chi_i^{[k-j]}t\right).
\end{equation*}
Particularly, we can know that the first three terms for $X_i^{[K]}$ are
\begin{equation*}
    \begin{split}
      X_i^{[1]}=&i \left(x+\chi_it\right) , \\
      X_i^{[2]}=&i \left[\chi_i^{[2]} x+\left(\chi_i^{[2]}\chi_i+\frac{1}{2}\right)t\right],\\
      X_i^{[3]}=&i \left[\chi_i^{[3]} x+\left(\chi_i^{[3]}\chi_i+\chi_i^{[2]} \right) t
 \right].
    \end{split}
\end{equation*}
Moreover, based on the elementary Schur polynomials we have the expansion
\begin{lem}\label{lem2}
\begin{equation*}
  \exp\left(\sum_{k=1}^{\infty}X_i^{[k]}\epsilon_i^{k}\right)=\sum_{j=0}^{\infty}S_i^{[j]}\epsilon_i^{j}
\end{equation*}
where $S_i^{[j]}$ is
\begin{equation*}
  S_i^{[j]}=\sum_{{\sum_{k=0}^{m}kl_k=j}}\frac{(X_i^{[1]})^{l_1}(X_i^{[2]})^{l_2}\cdots (X_i^{[m]})^{l_m}}{l_1!l_2!\cdots l_m!}.
\end{equation*}
\end{lem}
Specially
\begin{equation*}
    \begin{split}
    S_i^{[0]}=&1,\,\, S_i^{[1]}=X_i^{[1]},\\
      S_i^{[2]}=&\frac{1}{2}(X_{i}^{[1]})^{2}+X_{i}^{[1]},\\
      S_i^{[3]}=&X_{i}^{[3]}+X_{i}^{[1]}X_{i}^{[2]}+\frac{1}{6}(X_{i}^{[1]})^{3}.
    \end{split}
\end{equation*}
On the other hand, we need the following expansion series
\begin{lem}\label{lem3}
\begin{equation*}
  \frac{1}{\chi_i(\epsilon_i)\pm b_1}=\sum_{k=0}^{\infty}\mu^{[k]}_{i,\pm} \epsilon_i^{k}
\end{equation*}
where
\begin{equation*}
\begin{split}
   \mu^{[0]}_{i,\pm}&=\frac{1}{\left( {\pm b_1}+\chi_i \right)},\\
    \mu_{i,\pm}^{[k]}&=\frac{-1}{\left( {\pm b_1}+\chi_i \right)}\sum_{j=0}^{k-1}\mu_{i,\pm}^{[j]}\chi_i^{[k-j]},\,\, j\geq 1.
\end{split}
\end{equation*}
\end{lem}
Specially
\begin{equation*}
    \begin{split}
       \mu^{[1]}_{i,\pm}=&{\frac {-1}{ \left( {\pm b_1}+\chi_i \right) ^{2}}},\\
       \mu^{[2]}_{i,\pm}=&-{\frac {({\pm b_1}+\chi_i)\chi_i^{[2]}-1}{ \left( {\pm b_1}+\chi_i \right) ^{3}}},\\
\mu^{[3]}_{i,\pm}=&-{\frac { \left( {\pm b_1}+\chi_i \right) ^{2}\chi_i^{[3]}-2\left( {\pm b_1}+
\chi_i \right) \chi_i^{[2]}+1}{ \left( {\pm b_1}+\chi_i \right)^{4}}}.
    \end{split}
\end{equation*}
Finally, we have the asymptotical series for fundamental solution
\begin{equation}\label{asymp-sol}
    \Phi_i(\epsilon_i)=\sum_{j=0}^{\infty}\Phi_i^{[j]}\epsilon_i^j,
\end{equation}
where
\begin{equation*}
      \Phi_i^{[j]}=D\begin{bmatrix}
                    S_i^{[j]} \\
 {\displaystyle \sum_{m=0}^{j}\mu^{[m]}_{i,+}S_{i}^{[j-m]}}\\
                  {\displaystyle \sum_{m=0}^{j}\mu^{[m]}_{i,-}S_{i}^{[j-m]}}   \\
                  \end{bmatrix},
\end{equation*}which solve the following differential equations
\begin{equation}\label{lax-seed}
  \begin{split}
     \Phi_x=&U(\lambda_i+d_i\epsilon_i^2;Q[0])\Phi,  \\
     \Phi_t=&V(\lambda_i+d_i\epsilon_i^2;Q[0])\Phi,
  \end{split}
\end{equation}
where
\begin{equation*}
       Q[0]  =\begin{pmatrix}
        0 & ia_1\exp [-\theta_1] & ia_2\exp [-\theta_2] \\
        ia_1\exp [\theta_1] & 0 & 0 \\
        ia_2\exp [\theta_2] & 0 & 0 \\
      \end{pmatrix}
\end{equation*}
Using a simple symmetry $\epsilon_i\rightarrow -\epsilon_i$, we can obtain another asymptotical series $\Phi_i(-\epsilon_i)$ for fundamental solution. Based on the
solution $\Phi_i(\pm \epsilon_i)$, we can construct the high-order RW solution. We choose the asymptotical series as
\begin{equation}\label{special-solution}
\begin{split}
   \Theta_i\equiv\frac{\Phi_i(\epsilon_i)+\Phi_i(-\epsilon_i)}{2}&=\sum_{j=0}^{\infty}\Phi_i^{[2j]}\epsilon_i^{2j}, \\
   \Xi_i\equiv\frac{\Phi_i(\epsilon_i)-\Phi_i(-\epsilon_i)}{2\epsilon_i}&=\sum_{j=0}^{\infty}\Phi_i^{[2j+1]}\epsilon_i^{2j}.
\end{split}
\end{equation}
To introduce some free parameters for high-order RW solutions which can be used to vary the RW pattern distribution, we choose the special solution
\begin{equation*}
\begin{split}
    \widehat{\Psi_i}&=\Xi_i+\alpha_i(\epsilon_i)\Theta_i,  \\
    &=\left(\frac{\epsilon_i\alpha_i(\epsilon_i)+1}{2\epsilon_i}\right)\Phi_i(\epsilon_i)
    +\left(\frac{\epsilon_i\alpha_i(\epsilon_i)-1}{2\epsilon_i}\right)\Phi_i(-\epsilon_i)
\end{split}
\end{equation*}
where
\begin{equation*}
  \alpha_i(\epsilon_i)=\sum_{k=0}^{\infty}\alpha_i^{[k]}\epsilon_i^{2k}.
\end{equation*}
To represent the solution with a compact form, we replace the parameter $\epsilon_i\alpha_i(\epsilon_i)+1$ with
$\exp\left[\epsilon_i\alpha_i(\epsilon_i)\right].$
That is
\begin{equation*}
  \Psi_j(\epsilon_j)=\left(\frac{\exp\left[\epsilon_i\alpha_i(\epsilon_i)\right]\Phi_i(\epsilon_i)-
  \exp\left[-\epsilon_i\alpha_i(\epsilon_i)\right]\Phi_i(-\epsilon_i)}{2\epsilon_i}\right).
\end{equation*}
Moreover,
\begin{equation*}
  \Psi_j(\epsilon_j)=\sum_{m=1}^{\infty}\Psi_j^{[m-1]}\epsilon_j^{2(m-1)},\,\, \Psi_j^{[m-1]}=\Phi_j^{[2m-1]}+\alpha_j^{[m-1]}.
\end{equation*}
We expand
\begin{equation*}
    \frac{1}{\lambda_j(\epsilon_j)-\overline{\lambda_i(\epsilon_i)}}=\sum_{m=0}^{+\infty}\frac{(\overline{d_i}\overline{\epsilon_i}^2-d_j\epsilon_j^2)^{m}}{(\lambda_j-\overline{\lambda_i})^{m+1}},
\end{equation*}
and
\begin{equation*}
   \langle\Psi_i(\epsilon_i),\Psi_j(\epsilon_j)\rangle=\sum_{m=0}^{+\infty}\left[\sum_{n=0}^{m}\langle\Psi_i^{[m-n]},\Psi_j^{[n]}\rangle\epsilon_j^{2n}
   \overline{\epsilon_i}^{2(m-n)}\right],
\end{equation*}
where
\begin{equation*}
  \langle\Psi_i(\epsilon_i),\Psi_j(\epsilon_j)\rangle\equiv [\Psi_i(\epsilon_i)]^{\dag}\Psi_j(\epsilon_j).
\end{equation*}
By directly calculations, we can obtain that
\begin{widetext}
\begin{equation*}
\begin{split}
\frac{\langle\Psi_i(\epsilon_i),\Psi_j(\epsilon_j)\rangle}{\lambda_j(\epsilon_j)-\overline{\lambda_i(\epsilon_i)}}=&\left[\sum_{m=0}^{+\infty}\frac{(\overline{d_i}\overline{\epsilon_i}^2
    -d_j\epsilon_j^2)^{m}}{(\lambda_j-\overline{\lambda_i})^{m+1}}\right]\sum_{m=0}^{+\infty}\left[\sum_{n=0}^{m}\langle\Psi_i^{[m-n]},\Psi_j^{[n]}\rangle\epsilon_j^{2n}
   \overline{\epsilon_i}^{2(m-n)}\right]  \\
    =&\sum_{k=0}^{+\infty}\sum_{l=0}^{k}\left[\left(\sum_{m=0}^l\binom l m\frac{(\overline{d_i})^{l-m}(-d_j)^{m}}{(\lambda_j-\overline{\lambda_i})^{l+1}}\epsilon_j^{2m}\overline{\epsilon_i}^{2(l-m)}\right)
    \left(\sum_{n=0}^{k-l}\langle\Psi_i^{[k-l-n]},\Psi_j^{[n]}\rangle \epsilon_j^{2n}\overline{\epsilon_i}^{2(k-l-n)}\right)\right]\\
   =&\sum_{k=0}^{+\infty}\sum_{s=0}^{k}\left[\sum_{l=0}^{k}\sum_{m+n=s}^{\substack{0\leq m\leq l,\\0\leq n\leq k-l}}
    \binom l m \frac{(\overline{d_i})^{l-m}(-d_j)^{m}}{(\lambda_j-\overline{\lambda_i})^{l+1}}\langle\Psi_i^{[k-l-n]},\Psi_j^{[n]}\rangle\right]\epsilon_j^{2s}
   \overline{\epsilon_i}^{2(k-s)}
\end{split}
\end{equation*}
\end{widetext}
\begin{lem}\label{lem4}
Thus we can obtain
\begin{equation*}
    \frac{\langle\Psi_i(\epsilon_i),\Psi_j(\epsilon_j)\rangle}{\lambda_j(\epsilon_j)-\overline{\lambda_i(\epsilon_i)}}=
    \sum_{r,t=1}^{+\infty,+\infty}M^{[r,t]}_{i,j}\epsilon_j^{2(t-1)}\bar{\epsilon}_i^{2(r-1)},
\end{equation*}
where
\begin{equation*}
\begin{split}
  M^{[r,t]}_{i,j}&=\sum_{l=0}^{r+t-2}\sum_{m+n=t-1}^{\substack{0\leq n\leq r+t-l-2,\\
  0\leq m\leq l}}
    \binom l m \frac{(\overline{d_i})^{l-m}(-d_j)^{m}}{(\lambda_j-\overline{\lambda_i})^{l+1}}  \\
    & \langle\Psi_i^{[r+t-l-n-2]},\Psi_j^{[n]}\rangle
\end{split}
\end{equation*}
\end{lem}
Specially, we have
\begin{equation*}
    \begin{split}
       M^{[1,1]}_{i,j}=&\frac{\langle \Psi_i^{[0]}, \Psi_j^{[0]} \rangle }{\lambda_j-
\overline{\lambda_i}},\\
M^{[1,2]}_{i,j}=&{\frac {1}{\lambda_j-\overline{\lambda_i}} \left( \langle \Psi_i^{[0]}
,\Psi_j^{[1]} \rangle -{\frac { \langle \Psi_i^{[0]},\Psi_j^{[0]}
\rangle {d_j}}{\lambda_j-\overline{\lambda_i}}} \right) },\\
M^{[2,1]}_{i,j}=&{\frac {1}{\lambda_j-\overline{\lambda_i}} \left( \langle \Psi_i^{[1]},\Psi_j^{[0]}\rangle +{\frac{\langle\Psi_i^{[0]},\Psi_j^{[0]}
 \rangle {\it \overline{d_i}}}{\lambda_j-\overline{\lambda_i}}} \right) },
\end{split}
\end{equation*}
\begin{equation*}
    \begin{split}
M^{[2,2]}_{i,j}=& \left[-{\frac{\left(\langle \Psi_i^{[0]},
\Psi_j^{[0]}\rangle {d_j}+ \langle \Psi_i^{[0]},\Psi_j^{[1]}
 \rangle\left(-\lambda_j+\overline{\lambda_i} \right)  \right) {\overline{d_i}}}{
 \left( \lambda_j-\overline{\lambda_i} \right) ^{2}}}\right.\\&\left.+{\frac { \left( -\langle \Psi_i^{[0]},\Psi_j^{[0]} \rangle {\overline{d_i}}+
 \langle \Psi_i^{[1]},\Psi_j^{[0]} \rangle  \left( -\lambda_j+\overline{\lambda_i} \right)
 \right) {d_j}}{ \left( \lambda_j-\overline{\lambda_i} \right) ^{2}}}\right.\\&\left.+\langle \Psi
_i^{[1]},\Psi_j^{[1]} \rangle \right] \frac {1}{\lambda_j-\overline{\lambda_i}}.
    \end{split}
\end{equation*}
To obtain the high-order RW solution, we merely need to take limit $\epsilon_j\rightarrow 0.$
\begin{thm}
Therefore the general RW solution can be represented as
\begin{equation*}
\begin{split}
 Q[N]&=Q[0]+i[P,\sigma_3],\, P=-XM^{-1}X^{\dag},\\
M&=\begin{bmatrix}
     M_{1,1} & M_{1,2} \\
     M_{2,1} & M_{2,2} \\
   \end{bmatrix}
\end{split}
\end{equation*}
where
\begin{equation*}
    \begin{split}
    M_{1,1}&=\left(M_{1,1}^{[r,t]}\right)_{1\leq r,t\leq N_1},\,\,  M_{2,2}=\left(M_{2,2}^{[r,t]}\right)_{1\leq r,t\leq N_2},\\
    M_{1,2}&=\left(M_{1,2}^{[r,t]}\right)_{\substack{1\leq r\leq N_1;\\
    1\leq t\leq N_2}},\,\,  M_{1,2}=\left(M_{1,2}^{[r,t]}\right)_{\substack{1\leq r\leq N_2;\\
    1\leq t\leq N_1}},\\
      X& =\begin{bmatrix}
           X_1 \\
           X_2 \\
           X_3 \\
         \end{bmatrix}
=\left[\Psi_1^{[1]},\cdots,\Psi_1^{[N_1]},\Psi_2^{[1]},\cdots,\Psi_2^{[N_2]}\right].
    \end{split}
\end{equation*}
By simple linear algebra, we can represent above formula as
\begin{equation}\label{NRW}
    \begin{split}
      q_1[N]&=a_1\left[\frac{\det(M_1)}{\det(M)}\right]\exp{\theta_1},  \\
      q_2[N]&=a_2\left[\frac{\det(M_2)}{\det(M)}\right]\exp{\theta_2},
    \end{split}
\end{equation}
where
\begin{equation*}
    M_1=M-2Y_2^{\dag}X_1,\,\,M_2=M-2Y_3^{\dag}X_1,
\end{equation*}
 $Y_2=\frac{X_2{\rm e}^{-\theta_1}}{a_1}$ and
$Y_3=\frac{X_3{\rm e}^{-\theta_2}}{a_2}.$
\end{thm}
The generalized form can be used to derive rogue wave solution with arbitrary order without the constrain conditions on background fields and the differential with spectral parameter. The solution formula is given by a purely algebraic way. Especially, high-order RW with different fundamental patterns can be obtained, in contrast to the ones reported before \cite{lingzhao,ZZWZ,MQ}.

\section{Exact rogue wave solutions and their dynamics}
We find that many new patterns for RW excitation can exist in the coupled system without the constrain conditions before. In what following, we demonstrate the pattern dynamics of the first-order rogue wave solution, second-order rogue wave solution, and superposition of them respectively.

\textbf{a) Fundamental rogue wave}

The first-order RW solution can be given directly by the above formula. The solution can be presented as follows by some simplifications
\begin{eqnarray}
q_1[1]&=&
\left[1+\frac{-2{\rm i}r_1}{\chi_1+b_1}\frac{(x+p_1t)^2+r_1^2t^2+
\frac{{\rm i}(x+p_1t-{\rm i}r_1t)}{p_1+b_1+{\rm i}r_1}}{(x+p_1t+\frac{1}{2r_1})^2+r_1^2t^2+\frac{1}{4r_1^2}}\right]\nonumber\\
&& a_1 {\rm e}^{\theta_1},\\
q_2[1]&=&
\left[1+\frac{-2{\rm i}r_1}{\chi_1-b_1}\frac{(x+p_1t)^2+r_1^2t^2+
\frac{{\rm i}(x+p_1t-{\rm i}r_1t)}{p_1-b_1+{\rm i}r_1}}{(x+p_1t+\frac{1}{2r_1})^2+r_1^2t^2+\frac{1}{4r_1^2}}\right]\nonumber\\
&& a_2 {\rm e}^{\theta_2},
\end{eqnarray}
where $p_1=\mathrm{Re}(\chi_1)$, $r_1=\mathrm{Im}(\chi_1)$ and $\chi_1$ is a double root for cubic equation \eqref{cubic}.
We find that there are three different types of rogue wave solution:
\begin{itemize}
  \item If $\frac{(p_1\pm b_1)^2}{r_1^2}\geq 3$, then the rogue wave is called anti-eye-shaped rogue wave (or dark RW).
  \item If $\frac{1}{3}< \frac{(p_1\pm b_1)^2}{r_1^2}<3$, then the rogue wave is four-petaled rogue wave.
  \item If $\frac{(p_1\pm b_1)^2}{r_1^2}\leq\frac{1}{3}$, then the rogue wave is called eye-shaped rogue wave (or bright RW).
\end{itemize}
Similar patterns for RW have been demonstrated in \cite{Zhao4}. The explicit conditions for the RW transition were not given there. Additionally, breathers with these different pattern units were derived in a three-component coupled system \cite{Liu}. The value of $\frac{(p_1\pm b_1)^2}{r_1^2}$ can be used to make judgment on the RW pattern in the two components conveniently  based on the above results. Explicitly, $\frac{(p_1+ b_1)^2}{r_1^2}$ can be used to clarify the RW pattern in the first component, and $\frac{(p_1- b_1)^2}{r_1^2}$ is used to clarify RW pattern for the second component.  When the multi RWs interact with each other, there will be many different patterns for which it is hard to know which fundamental RWs constitute them. These criterions can be used to clarify the fundamental RW pattern conveniently.

For the general case, since the roots for cubic equation and quartic equation are presented with the radical solution very complexly, a simple way to replace these radical solution is using the numerical roots. The above forms can be used to investigate dynamics of them directly.  Especially, we  can investigate RW solution analytically with $a_1=a_2$ case under which the RW pattern can be clarified explicitly. Since CNLS possesses the scaling symmetry, we can set $a_1=a_2=1$ without losing generality in this case. Setting $b_1=\frac{1}{2}\sqrt{1-\gamma_1^2}$, then we can have
\begin{equation*}
    \begin{split}
       \lambda_1=&\frac{i\delta_{+}(3+\gamma_1)}{4(1+\gamma_1)},\,\, \chi_1=\frac{i\delta_{+}}{2}  \\
       \lambda_2=&\frac{i\delta_{-}(3-\gamma_1)}{4(1-\gamma_1)},\,\, \chi_2=\frac{i\delta_{-}}{2}
    \end{split}
\end{equation*}
where
\begin{equation*}
    \delta_{\pm}=\sqrt{(1\pm \gamma_1)(3\pm \gamma_1)}.
\end{equation*}
If $0<b_1<1/2$,  we can obtain the RW with velocity equals to zero. RW patterns in the two components are identical with each other in this case. And the types of RW are determined by
the value $\frac{1-\gamma_1}{3+\gamma_1}$ and $\frac{1+\gamma_1}{3-\gamma_1}$ respectively. It is readily to see that there can be both eye-shaped or four-petaled RW exists in the two components for this case. It is impossible for anti-eye-shaped RW to exist correspondingly in both components for the coupled model, since  $\frac{1-\gamma_1}{3+\gamma_1}$ and $\frac{1+\gamma_1}{3-\gamma_1}$ can not be larger than 2 for $0<b_1<1/2$ (namely $|\gamma_1|< 1$). However, the anti-eye-shaped RW can exist in both components for the coupled model with negative nonlinear terms \cite{Li}.

\begin{figure}[htb]
\centering
\subfigure[$|q_1|^2$]{\includegraphics[height=32mm,width=38.5mm]{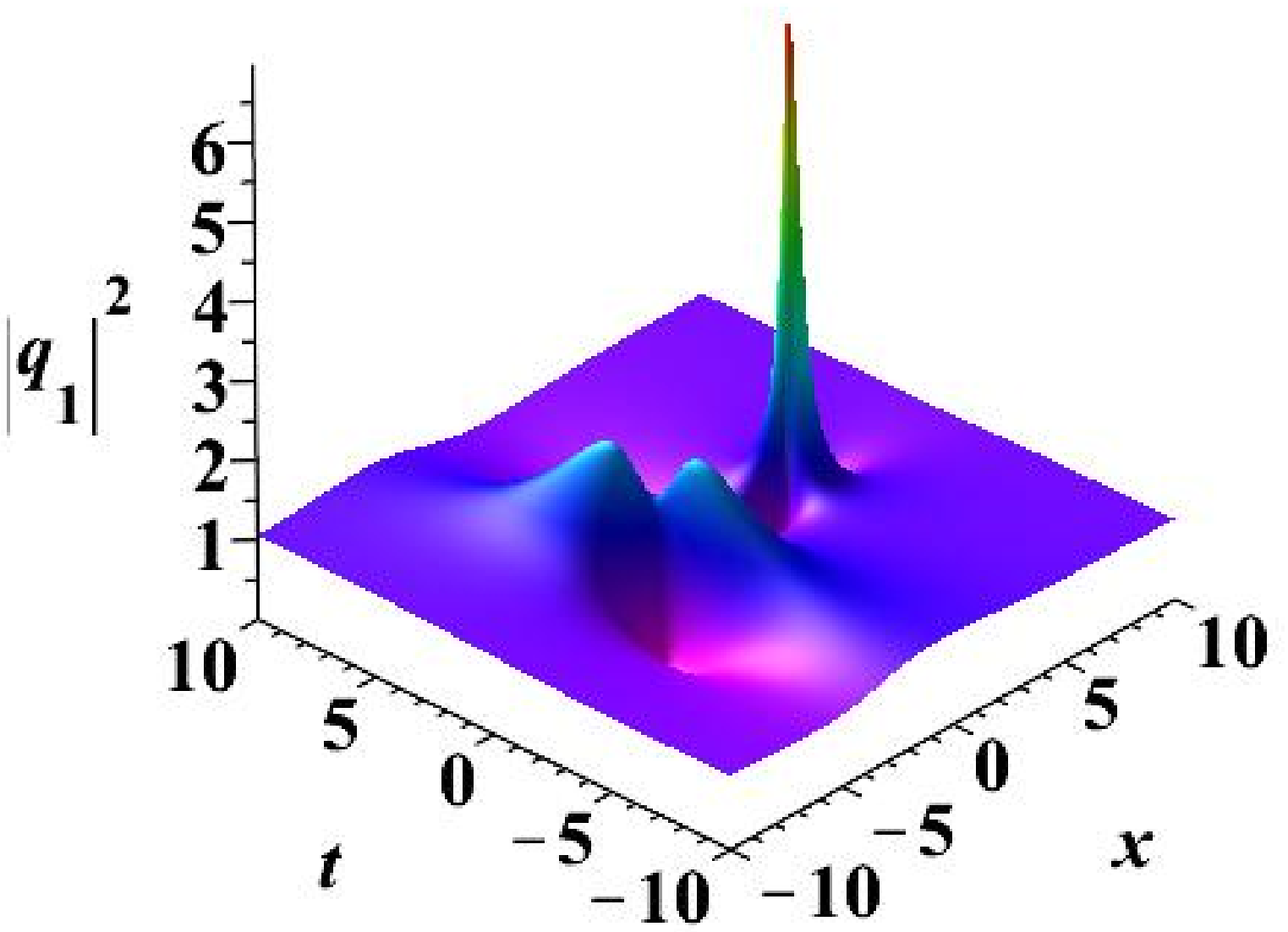}}
\hfil
\subfigure[$|q_2|^2$]{\includegraphics[height=32mm,width=38.5mm]{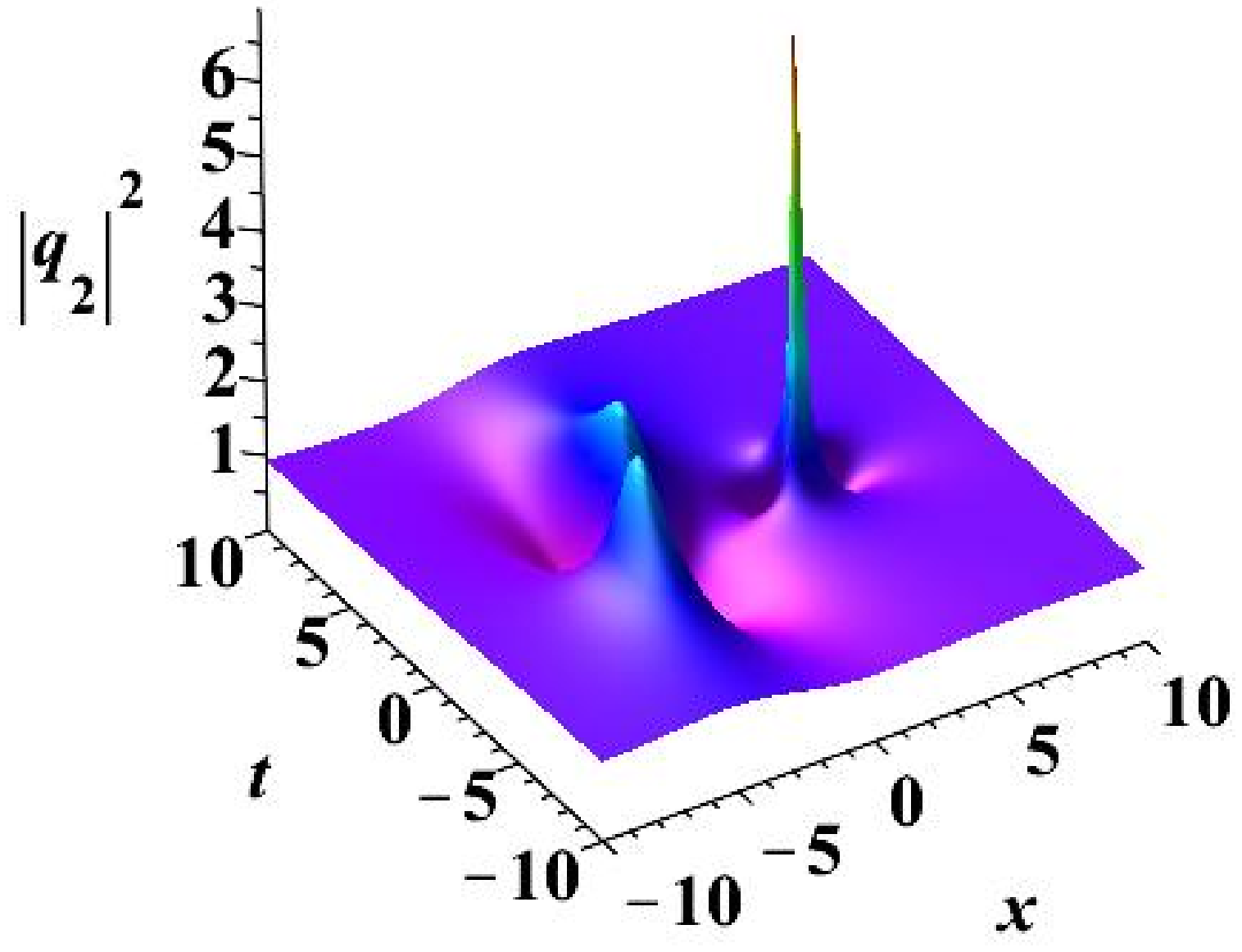}}
\caption{(color online): The coexistence of eye-shaped rogue wave and four-petaled one in each component.  Parameters: $b_1=\frac{2}{5}$,$\gamma_1=\frac{3}{5}$, $a_1=a_2=1$, $\alpha_1^{[0]}=\frac{17}{5}$, $\alpha_2^{[0]}=\frac{\sqrt{6}}{5}-1$ and $N_1=N_2=1$.}\label{fig1}
\end{figure}

If $b_1>1/2$, then we introduce
\begin{equation*}
    \begin{split}
        k_1&={\frac{4\kappa}{1+\kappa}}+{\frac {i\sqrt {\kappa({\kappa}^{2}-14\,\kappa+1)}}{1+\kappa}},  \\
        \gamma_1&=\frac{1}{2}(k_1+\frac{1}{k_1})-2,\,\, b_1=\frac{\kappa^2-6\kappa+1}{4(1+\kappa)\sqrt{\kappa}}\\
       \delta_+&=\frac{1}{2}(k_1-\frac{1}{k_1}),\,\, \delta_-=\overline{\delta_+} \\
    \end{split}
\end{equation*}
where $\kappa\in (0,7-4\sqrt{3})\cup (7+4\sqrt{3},\infty).$
Furthermore, we have
\begin{equation}\label{type1}
\begin{split}
    p_1&=-{\frac {\sqrt{{\kappa}^{2}-14\kappa+1}}{4\sqrt {\kappa}}},\,\, p_2=-p_1  \\
    r_1&=\frac{\kappa-1}{1+\kappa},\,\, r_2=r_1.
\end{split}
\end{equation}
We can verify that $\frac{(p_1+b_1)^2}{r_1^2}<1/3$ for eye-shaped RW and $\frac{(p_1-b_1)^2}{r_1^2}>1/3$ for four-petaled RW or anti-eye-shaped one.  And the types for RW pattern in $q_1$ and $q_2$ components are different.

\textbf{b) Superposition for fundamental rogue wave solutions}

Superposition for fundamental RW solutions here is refer to nonlinear superposition for two fundamental RW solutions with different spectral parameter setting for the double root $\chi$. This will make it possible to obtain RW with different patterns coexisting, since the RW pattern is determined by the value of $\chi$ and relative wave vector parameter $b_1$. It has been shown above that the RW pattern for two components can be clarified to two cases for different $b_1$ values.
Firstly, we give two special cases which show that there are two different types of dynamics behavior. Similar coexistence of RW with different patterns were also demonstrated in other coupled system with some certain constrains on background fields \cite{Liu}.  It should be emphasized that there is no constrains on background field here. We show that eye-shaped RW can coexist with a four-petaled one
 or an anti-eye-shaped one. Nextly, we discuss them explicitly.

When  $0<b_1<1/2$ and $a_1=a_2=1$, we can obtain the two RWs with different patterns in each component. One pattern is a four-petaled structure, and the other is an eye-shaped structure.  The types for two RWs in components $q_1$ and $q_2$ are consistence  (Fig. \ref{fig1}). In this case, it is impossible to obtain an anti-eye-shaped RW coexist with one eye-shaped RW, since anti-eye-shaped RW can not exist correspondingly in both components for the coupled model as discussed above.

\begin{figure}[htb]
\centering
\subfigure[$|q_1|^2$]{\includegraphics[height=32mm,width=38.5mm]{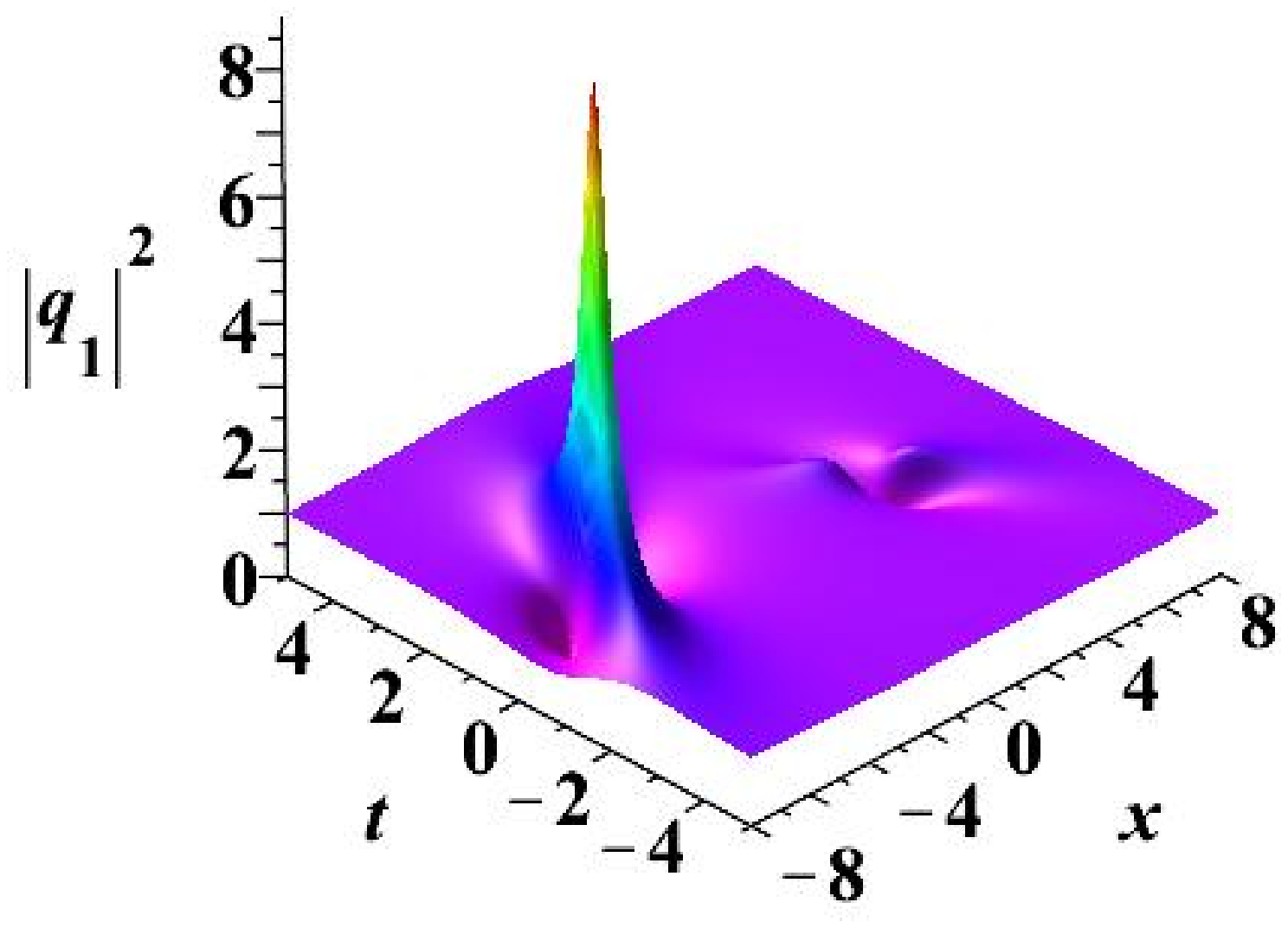}}
\hfil
\subfigure[$|q_2|^2$]{\includegraphics[height=32mm,width=38.5mm]{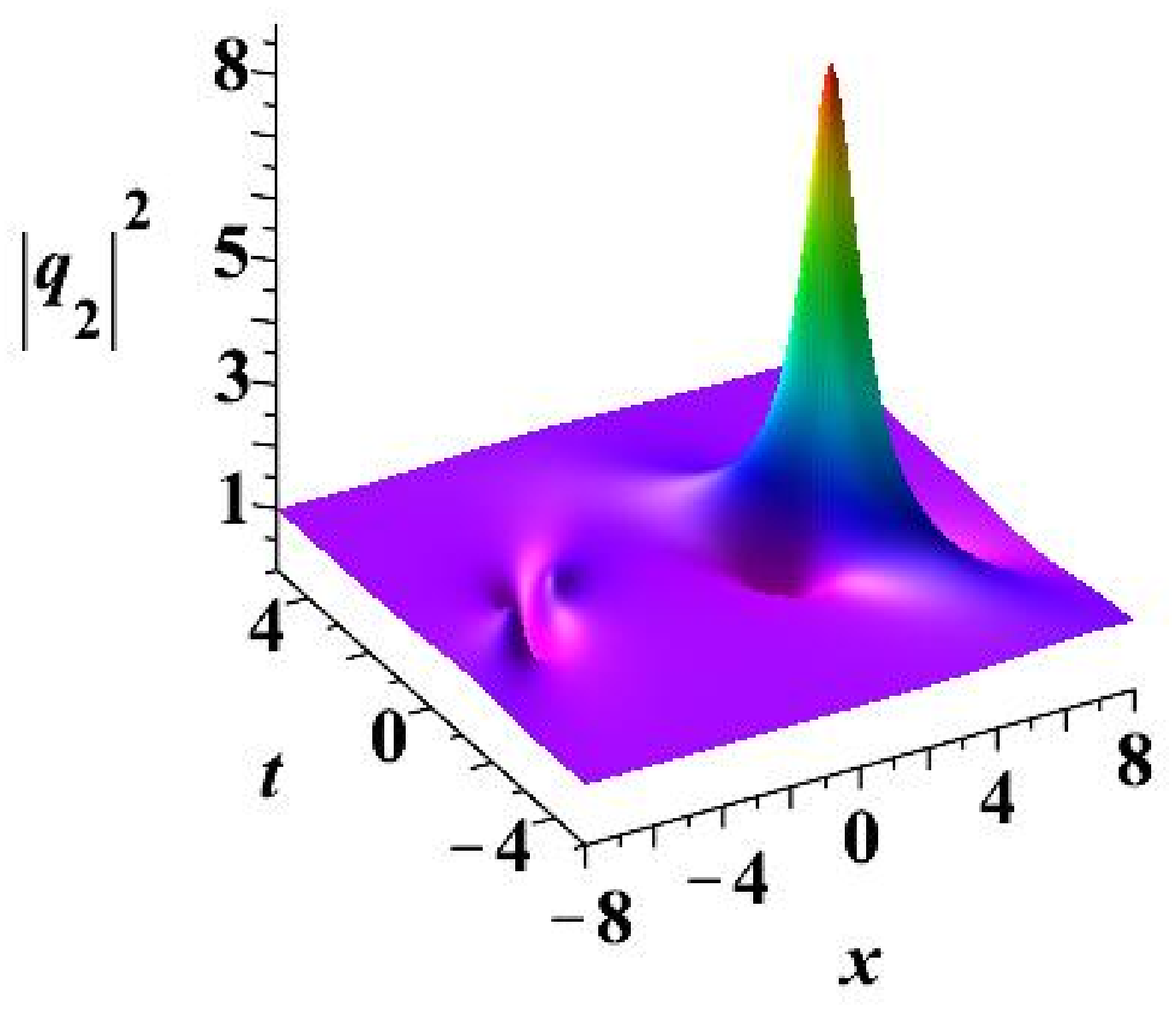}}
\caption{(color online): The coexistence of eye-shaped rogue wave and anti-eye-shaped one in each component. Parameters: $\kappa=\frac{1}{25}$, $b_1=\frac{119}{130}$, $a_1=a_2=1$, $\alpha_1^{[0]}=-\frac{76}{13}$, $\alpha_2^{[0]}=\frac{28}{13}$, $N_1=N_2=1$.}\label{fig2}
\end{figure}

When  $b_1>1/2$ and $a_1=a_2=1$, we can obtain the two RW solution which admits one anti-eye-shaped RW or a four-petaled RW and an eye-shaped one in each component. The types for two rogue waves in components $q_1$ and $q_2$ are different. As an example, we show one case for anti-eye-shaped RW coexisting with eye-shaped RW in Fig. \ref{fig2}. It is seen that at the same location, one anti-eye-shaped RW in one component corresponds to an eye-shaped RW in the other component.  It should be noted that the case for a four-petaled RW coexisting with an eye-shaped RW is distinctive from the ones in Fig. \ref{fig1}.

\begin{figure}[htb]
\centering
\subfigure[$|q_1|^2$]{\includegraphics[height=32mm,width=38.5mm]{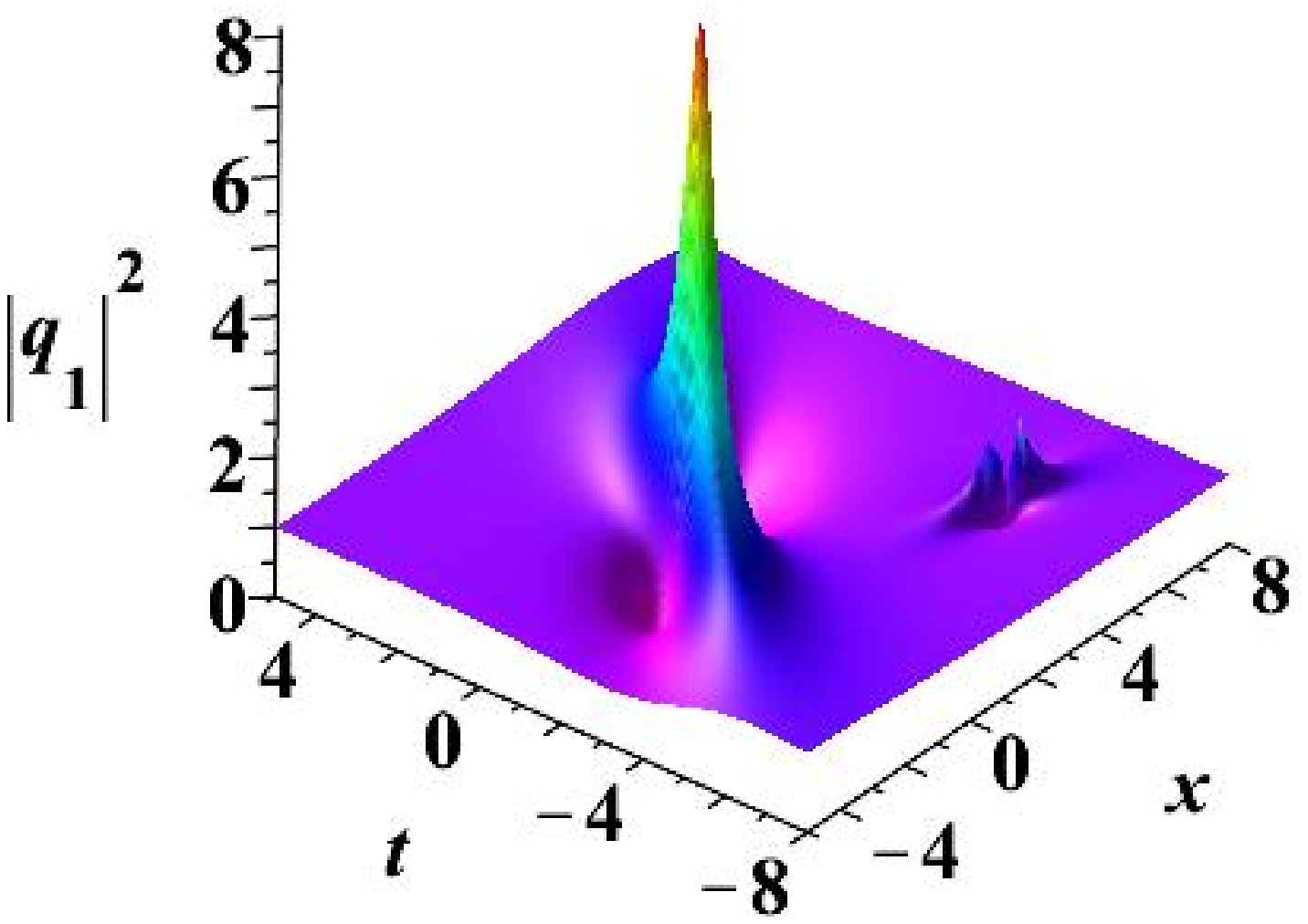}}
\hfil
\subfigure[$|q_2|^2$]{\includegraphics[height=32mm,width=38.5mm]{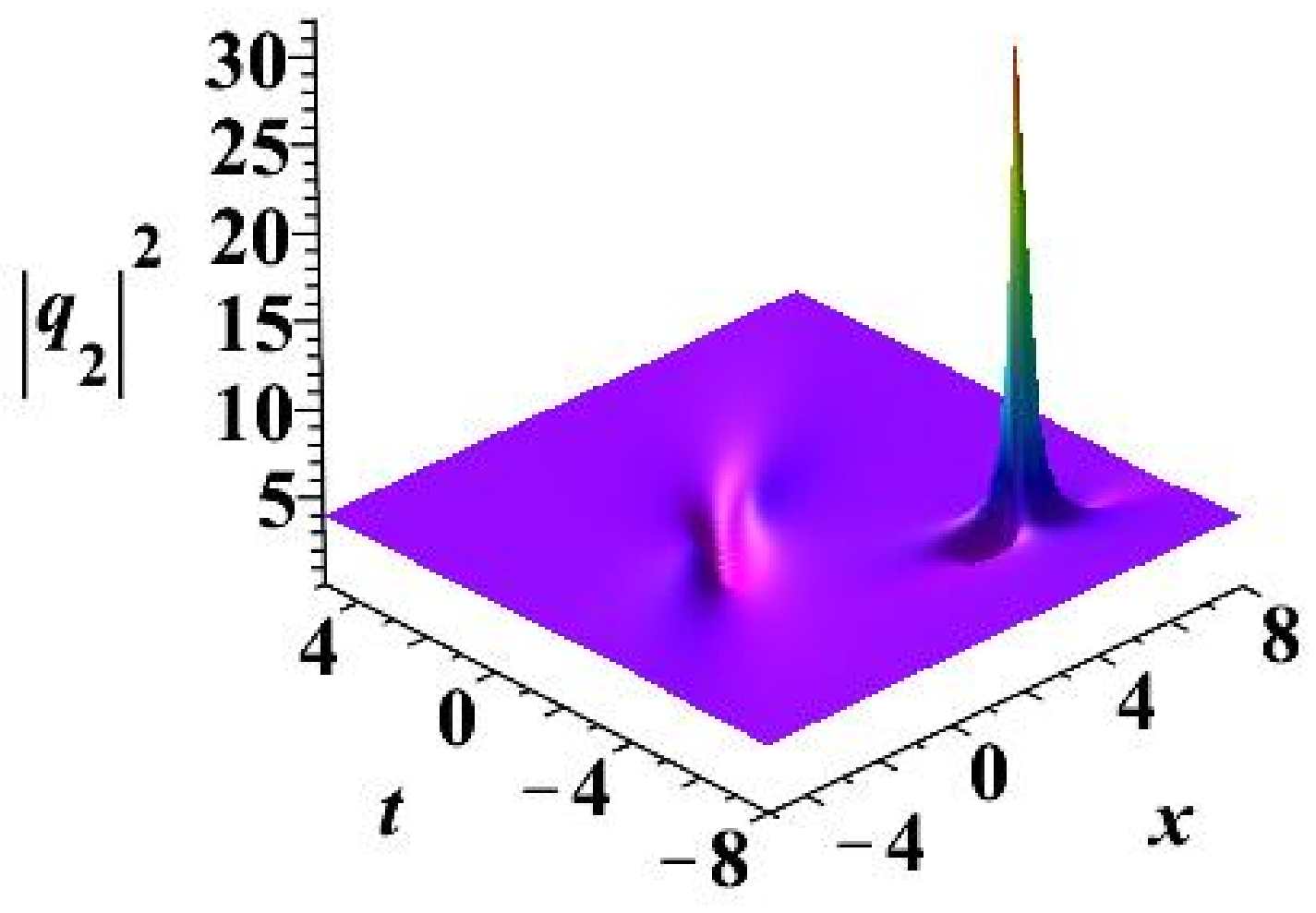}}
\caption{(color online): The coexistence of eye-shaped rogue wave and four-petaled one in component $q_1$, and eye-shaped rogue wave with an anti-eye-shaped one in component $q_2$. $1/3<\frac{(p_1+b_1)^2}{r_1^2}\approx 0.8744365601<3$ four-petaled rogue wave, $0<\frac{(p_1-b_1)^2}{r_1^2}\approx 0.0045075247<1/3$ bright rogue wave,
$\frac{(p_2+b_1)^2}{r_2^2}\approx 0.03465434964<1/3$ bright rogue wave, $\frac{(p_2-b_1)^2}{r_2^2}\approx 6.722765218>3$ dark rogue wave}\label{fig3}
\end{figure}

For the general case $a_1 \neq a_2$, we merely give a special example to show its dynamics. For instance, we choose the parameters
\begin{equation*}
  \begin{split}
   b_1=&1,\,\, a_1=1,a_2=2,\,\, N_1=N_2=1,\\
     \lambda_1=&\frac{3-i\sqrt {207+48\sqrt {3}}}{8},\,\, \lambda_2=\frac{3-i\sqrt {207-48\sqrt {3}}}{8},  \\
     \chi_1=&\frac{\sqrt {3}}{2}-i{\frac {6+\sqrt {3}}{66}}\sqrt {207+48\,
\sqrt {3}},\,\, \alpha_1^{[0]}=10,\\ \chi_2=&-\frac{\sqrt {3}}{2}+i{\frac {\sqrt {3}-6}{66}}\sqrt {207-48
\sqrt {3}},\,\, \alpha_2^{[0]}=0,
  \end{split}
\end{equation*}
then we obtain the figure (Fig. \ref{fig3}). One eye-shaped RW and a four-petaled RW coexist in one component, and an anti-eye-shaped RW with an eye-shaped RW coexist in the other component. These would provide us more interesting excitation patterns for two RW case in coupled system. It should be noted that the above patterns all contains eye-shaped RW. Can an anti-eye-shaped RW coexist with a four-petaled one? It has been shown that this pattern can emerge in one component in a three-component coupled system \cite{Liu}. We are not sure whether this pattern can emerge in this two-component coupled system.

\textbf{c) Second-order rogue wave solution}

As the second-order RW in scalar system \cite{Yang,Ling,He,Ling2,Akhmediev}, the second-order RW solution here is still obtained by superposition of fundamental RW solutions with the same spectral parameter. Therefore, one can obtain three fundamental RWs with identical pattern in each component for the second-order RW solution, which is similar to the ones in scalar case. But the three fundamental RWs can be three eye-shaped ones, three anti-eye-shaped ones, and three four-petaled ones, in contrast three eye-shaped ones in scalar system.  By choosing parameters, we can obtain the different types of RW solution. The three RWs can be superposed together, and construct symmetric structure as the second-order RW with highest peak for scalar NLS \cite{Yang,Ling,He,Ling2,Akhmediev}. However, it is usually very complicated to obtain the symmetric structure since there are much more parameters than the one for scalar NLS. Based on the solution form presented here, we can obtain them more easily. We show them by three categories.

\begin{figure}[htb]
\centering
\subfigure[$|q_1|^2$]{\includegraphics[height=32mm,width=38.5mm]{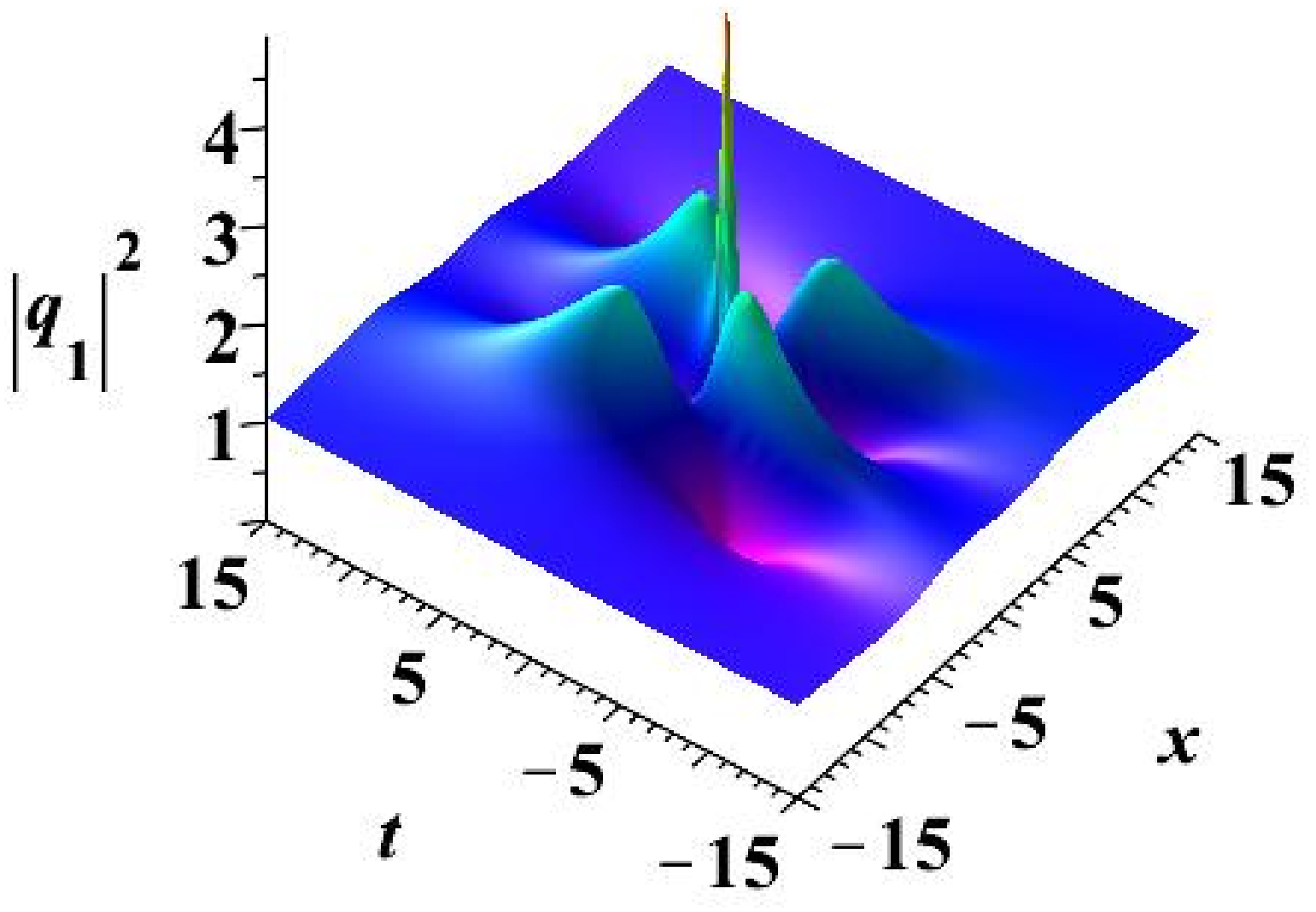}}
\hfil
\subfigure[$|q_2|^2$]{\includegraphics[height=32mm,width=38.5mm]{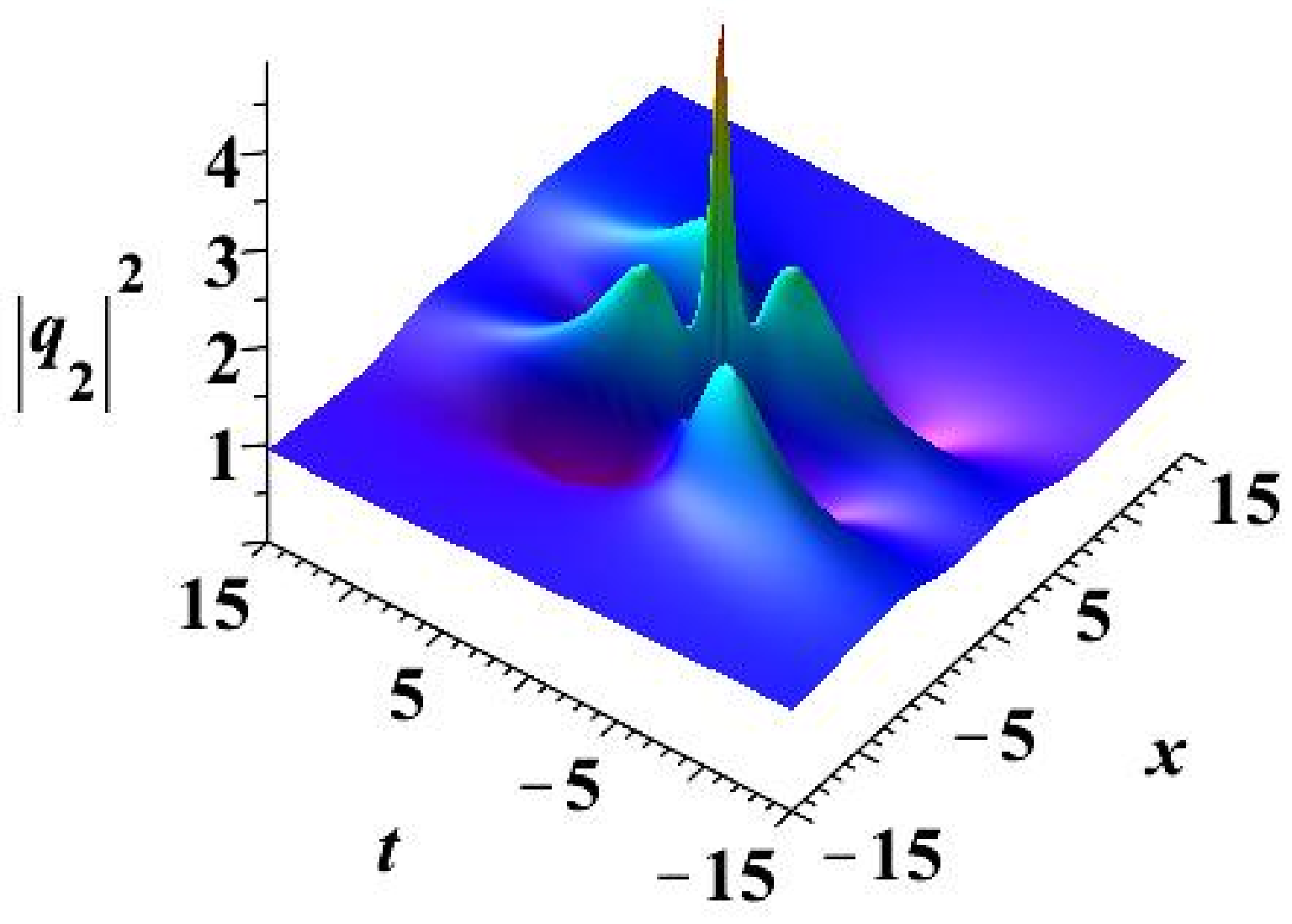}}
\caption{(color online): The superposition of three four-petaled rogue waves in both components for the second-order rogue wave solution. The parameters are: $b_1=\frac{2}{5}$, $a_1=a_2=1$, $\lambda_1=\frac{3i}{5}\sqrt{6}$, $\chi_1=\frac{i}{5}\sqrt{6}$, $\alpha_1^{[0]}=0$, $\alpha_1^{[1]}=0$, $N_1=2$, $N_2=0$.}\label{fig4}
\end{figure}

With $b_1<1/2$ and $a_1=a_2=1$, it is possible to obtain three eye-shaped RWs or four-petaled RWs in both components. As an example, we show dynamics of the second-order RW with highest peak with $b_1=2/5$ in (Fig. \ref{fig4}). Then, which three fundamental RW pattern constitute the pattern in Fig. 4?
The values of parameters $b_1$ and $\chi$ can be used to analyze the patterns in Fig. 4 based on the above criterions for clarifying fundamental RW.  It is proven that the patterns in the two components are both superimposed by three four-petaled RWs.
\begin{figure}[htb]
\centering
\subfigure[$|q_1|^2$]{\includegraphics[height=32mm,width=38.5mm]{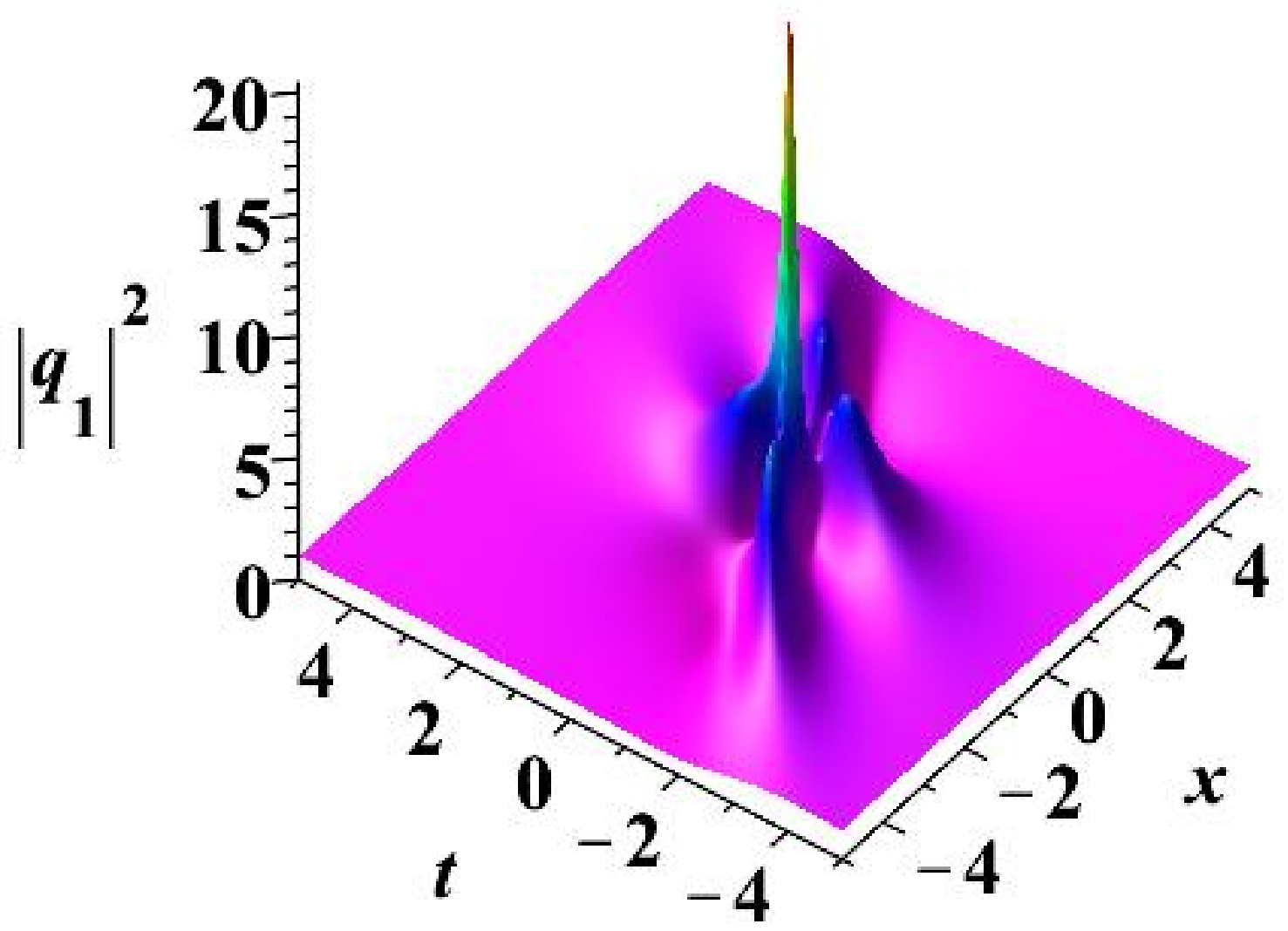}}
\hfil
\subfigure[$|q_2|^2$]{\includegraphics[height=32mm,width=38.5mm]{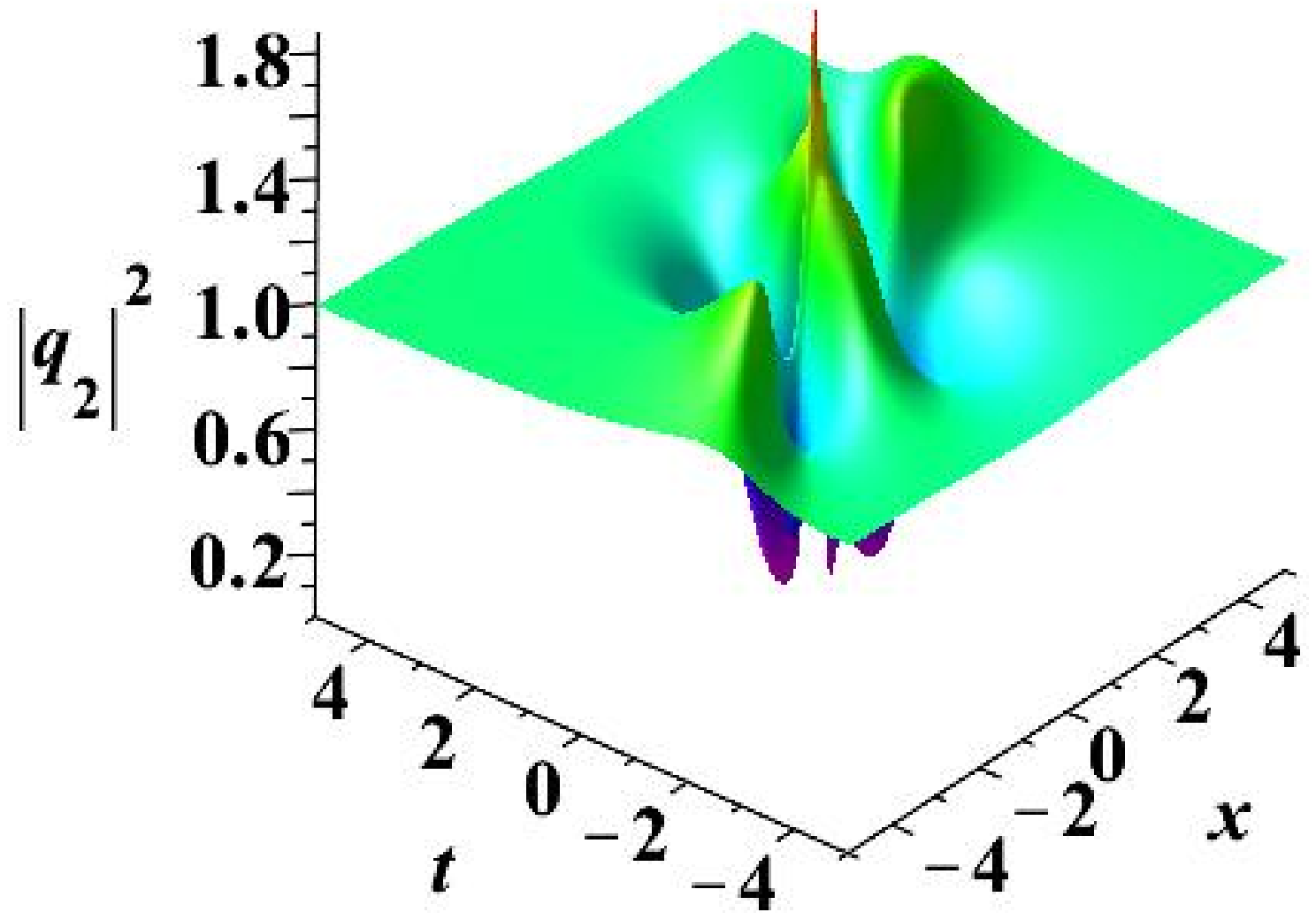}}
\caption{(color online): The superposition of three eye-shaped rogue waves in the first component, and three anti-eye-shaped rogue waves in the second component. The parameters are: $b_1=\frac{119}{130}$, $\lambda_1=-{\frac {69}{2380}}\,\sqrt {69}-{\frac {1728}{1547}}\,i$, $\chi_1=-{\frac {12}{13}}i-\frac{1}{10}\sqrt{69}$, $\alpha_1^{[0]}=0$, $\alpha_1^{[1]}=0$,  $N_1=2$, $N_2=0$.}\label{fig5}
\end{figure}

The second case is with $b_1>1/2$ and $a_1=a_2=1$. In this case, the patterns for RW in two components are different. As an example, we show the dynamics of them with $b_1=119/130$ in Fig. \ref{fig5}. It is seen that the patterns in two components are indeed different.  Then, we can also clarify which fundamental RW pattern superpose the patterns in Fig. \ref{fig5}. For the patterns in Fig. 5, $b_1=\frac{119}{130}$, $p_1= -\frac{\sqrt{69}}{10}$, and $r_1=-\frac {12}{13}$. From the value of $\frac{(p_1\pm b_1)^2}{r_1^2}$, we can know that the pattern in $q_1$ is superimposed by three eye-shaped RWs, and the pattern in $q_2$ is constituted of three anti-eye-shaped ones. It is shown that three eye-shaped RWs admits the largest peak value among these three different pattern superposition ways, and the anti-eye-shaped ones admits the lowest value (compare the peak values in Fig. 4 and 5).

The third case is the general case with $a_1\neq a_2$, $b_1\neq0$. The dynamics of RW can be investigated by second-order solution. The patterns of them are superposed by three eye-shaped ones (similar to Fig. 5(a)), or three four-petaled ones (similar to Fig. 4(a) or (b)), or three anti-eye-shaped ones (similar to Fig. 5(b)). Therefore, we do not show the figures.

\begin{figure}[htb]
\centering
\subfigure[$|q_1|^2$]{\includegraphics[height=32mm,width=38.5mm]{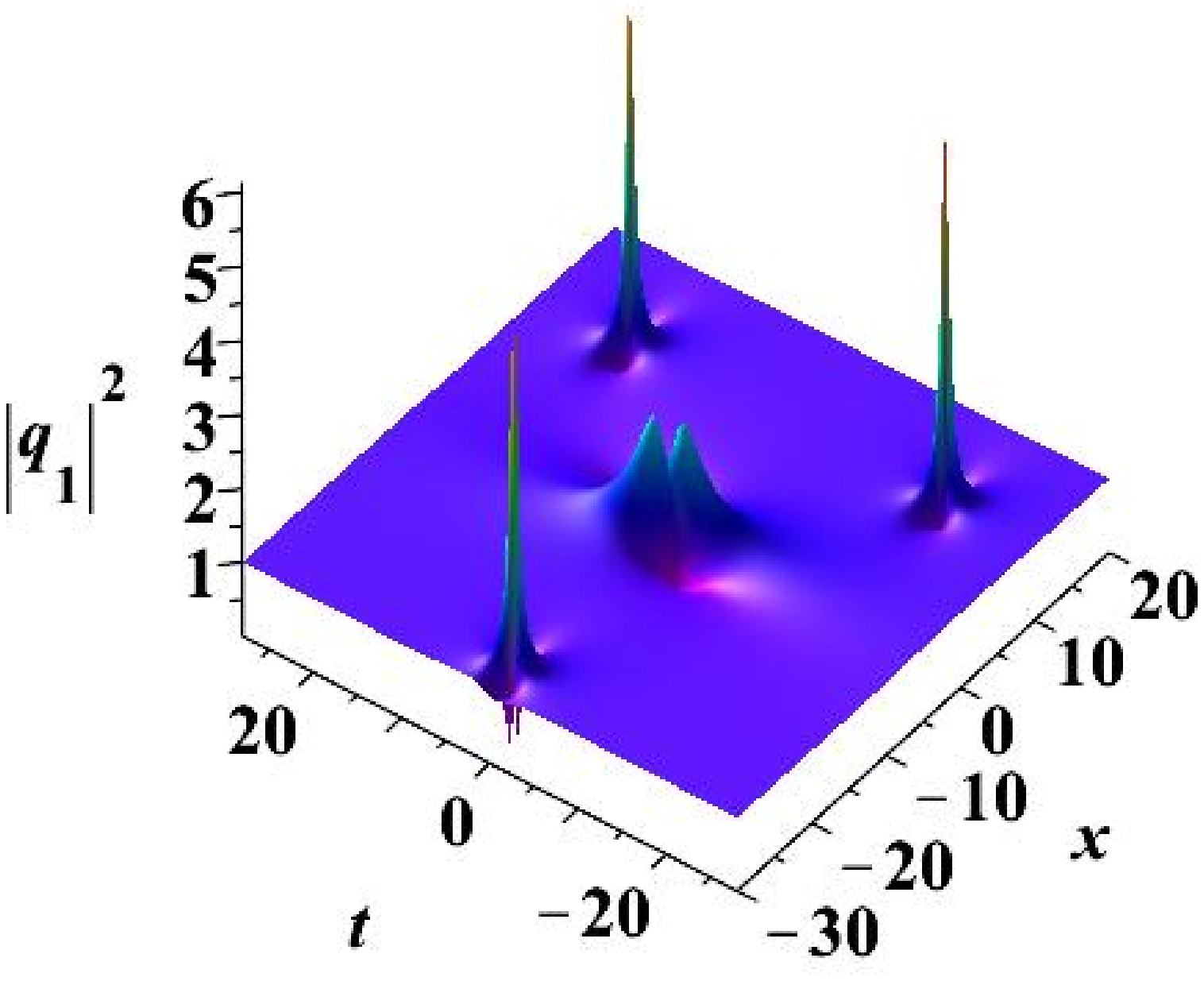}}
\hfil
\subfigure[$|q_2|^2$]{\includegraphics[height=32mm,width=38.5mm]{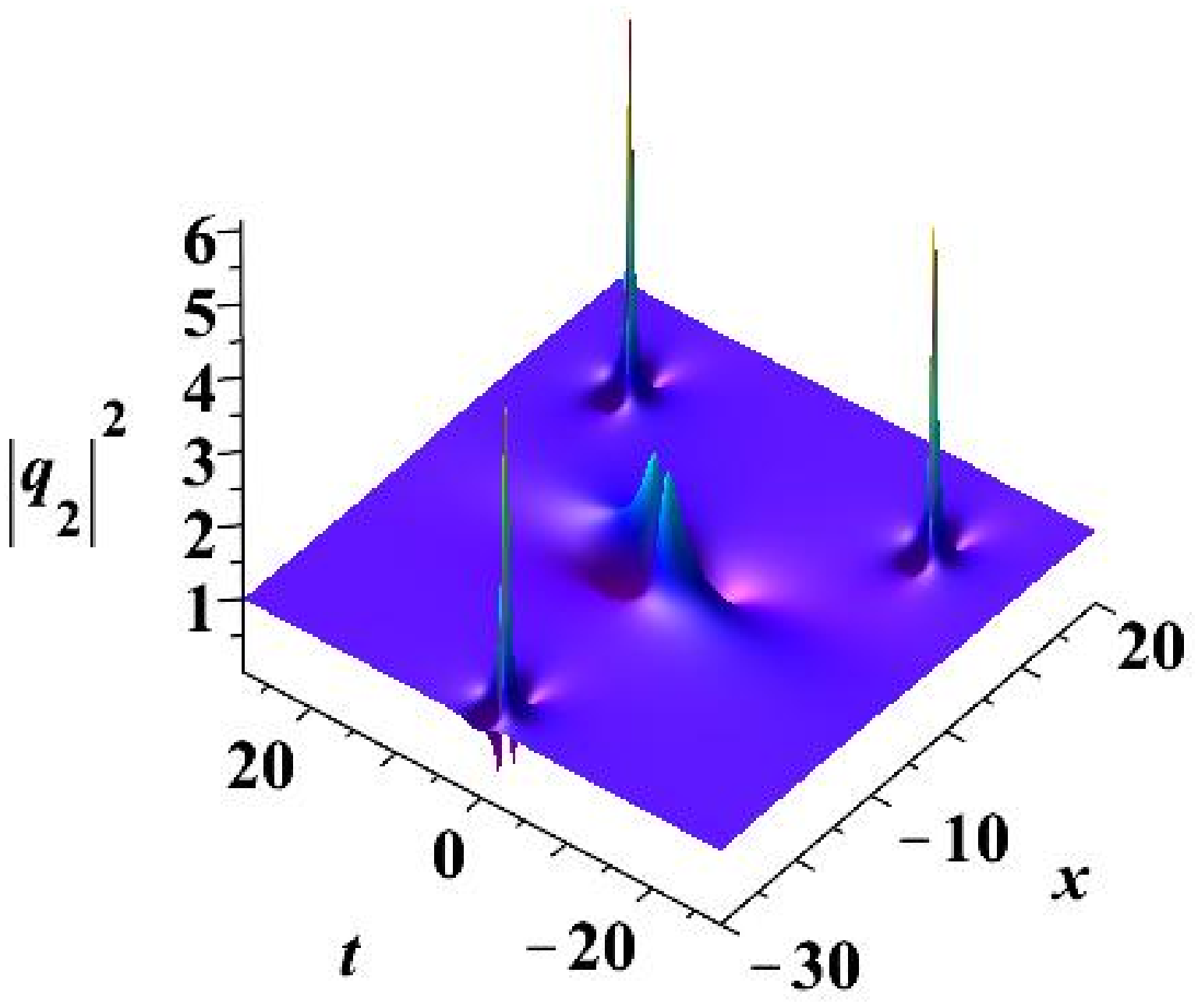}}
\caption{(color online): The  coexistence of one four-petaled RW and three eye-shaped ones. Parameters: $b_1=\frac{2}{5}$, $a_1=a_2=1$, $\lambda_1=\frac{27i}{20}$, $\lambda_2=\frac{3\sqrt{6}i}{5}$, $\chi_1=\frac{6i}{5}$, $\chi_2=\frac{\sqrt{6}i}{5}$, $\alpha_1^{[0]}=\frac{12}{5}i$, $\alpha_1^{[1]}=5000i$, $\alpha_2^{[0]}=\frac{2\sqrt{6}}{5}i$,
$N_1=2$ and $N_2=1$.}\label{fig6}
\end{figure}

\textbf{d) Superposition for the second-order rogue wave solution}

From the superposition of these different types RW solutions, we can obtain some novel dynamics behavior. We give them by two different categories. The first case is the interaction between the first-order RW  and second-order RW. The second case is the interaction between two second-order RWS.

The superposition for the interaction between the first-order RW and second-order RW can demonstrate cases that there are four RWs on the temporal-spatial distribution plane. Especially, the four RWs can admit many different patterns. It is possible to obtain at least $2+2+2+2=8$ different patterns, such as three eye-shaped ones with one anti-eye-shaped RW in both components, three eye-shaped ones with a four-petaled RW in both components(which includes two distinctive cases: RW patterns in the two components at the same locations are identical or different), the combined ones in each component(corresponding to Fig. 3 case), and the inverse cases of them.  For example, we show one case that one four-petaled RW coexist with three eye-shaped RWs in Fig. \ref{fig6}. All these patterns are in contrast to the four eye-shaped ones obtained before \cite{lingzhao}. The nonlinear interactions among them can induce many other different patterns based on these 8 fundamental pattern combinations.

\begin{figure}[htb]
\centering
\subfigure[$|q_1|^2$]{\includegraphics[height=32mm,width=38.5mm]{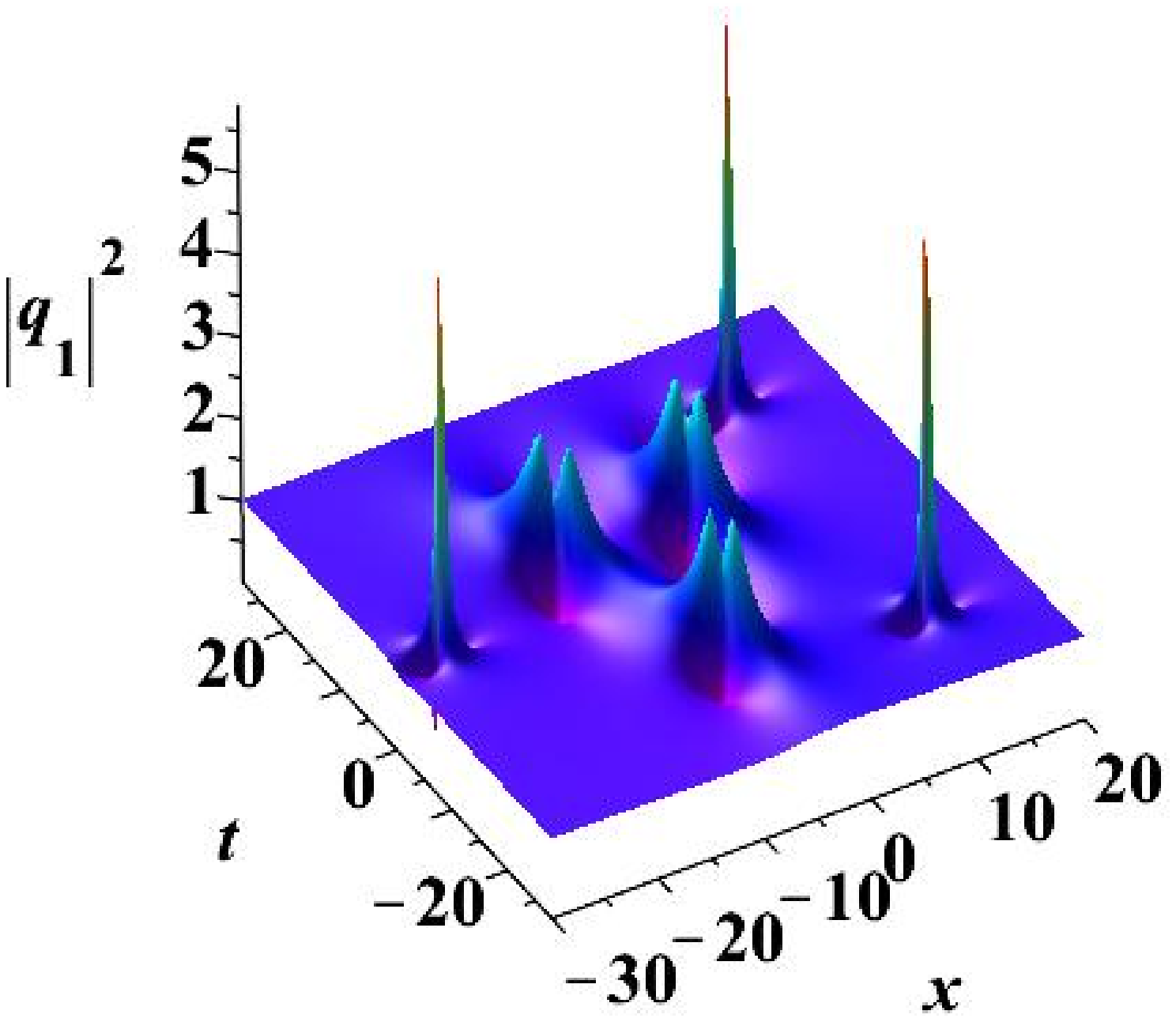}}
\hfil
\subfigure[$|q_2|^2$]{\includegraphics[height=32mm,width=38.5mm]{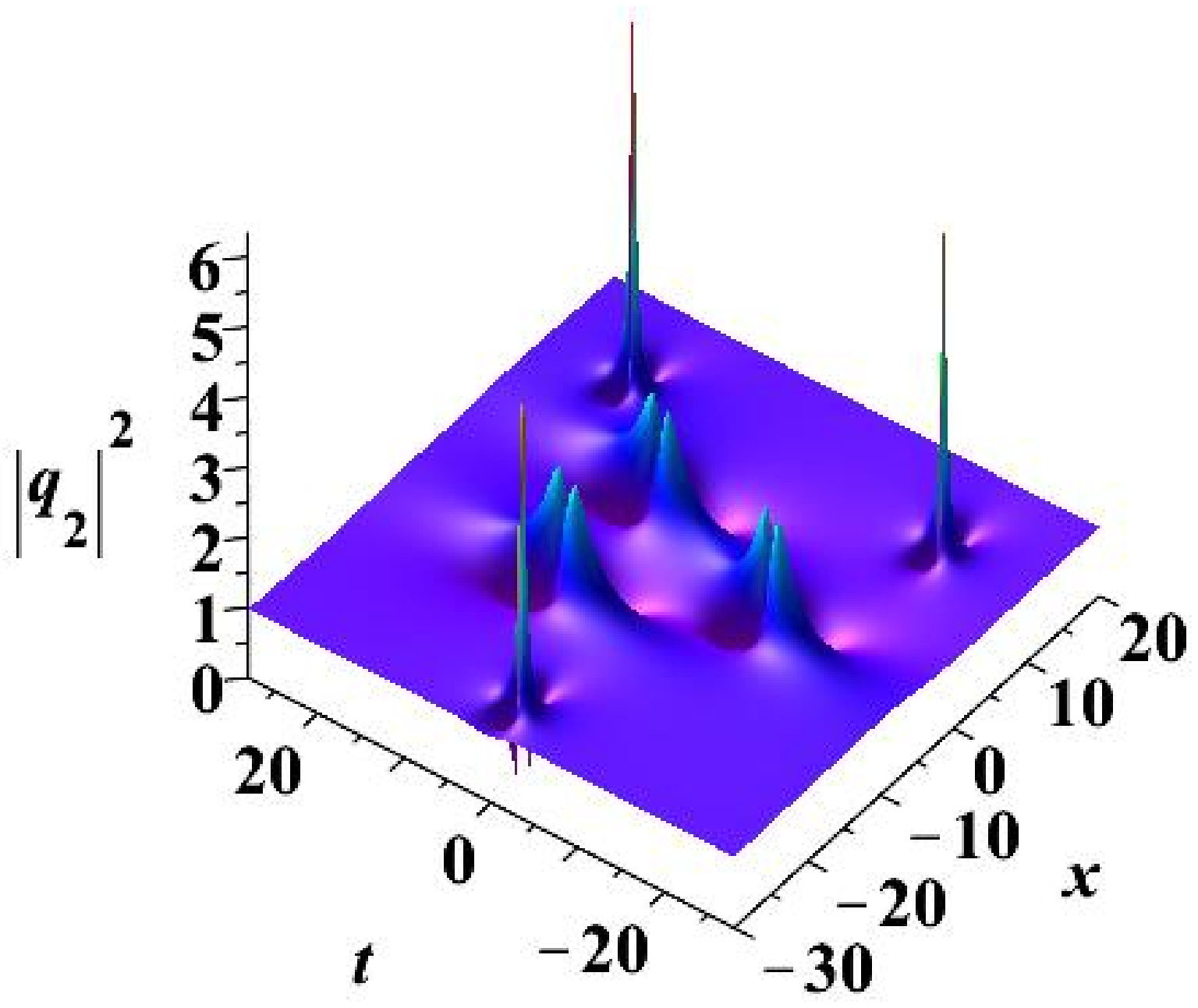}}
\caption{(color online): The coexistence of three four-petaled RW and three eye-shaped ones. Parameters: $b_1=\frac{2}{5}$, $a_1=a_2=1$,  $\lambda_1=\frac{27i}{20}$, $\lambda_2=\frac{3\sqrt{6}i}{5}$, $\chi_1=\frac{6i}{5}$, $\chi_2=\frac{\sqrt{6}i}{5}$, $\alpha_1^{[0]}=\frac{12}{5}i$, $\alpha_1^{[1]}=5000i$, $\alpha_2^{[0]}=\frac{2\sqrt{6}}{5}i$, $\alpha_2^{[1]}=100$, $N_1=N_2=2$.}\label{fig7}
\end{figure}

Moreover, we can observe the superposition of two second-order RW solutions. There are six RWs in the distribution plane. The patterns of them can be different, in contrast to the six eye-shaped ones obtained with constrain conditions on background fields \cite{lingzhao}.  The six RWs just possess two types of fundamental RW pattern and each type pattern is occupied by three RWs. We find that it is possible to obtain at least $4$ different patterns, mainly including three cases corresponding to Fig. 1-3 and one case similar to Fig. 1 but with RW patterns in the two components at the same locations are different.  As an example, we show one case that three four-petaled RWs and three eye-shaped ones coexist in Fig. \ref{fig7}. Many different superposition patterns can be obtained by varying the parameters based on these 4 fundamental pattern combinations.

\section{Conclusion}

In this paper, we present a method to derive high-order RW for coupled NLS model with no constrain conditions on background fields. It is demonstrated that coexistence of RWs with different  patterns can emerge in the coupled system.  For example, one four-petaled RW and three eye-shaped ones can coexist, in contrast to the four eye-shaped ones reported before. Three four-petaled RW and thee eye-shaped ones can constitute some new patterns for six fundamental ones case. There are many other different cases for the coexistence of RWs with different patterns. The superposition of different order RW solution can demonstrate more complex dynamics for RW excitation in the coupled systems. These RW pattern excitations are all in contrast to the ones in scalar NLS equation and the previous results obtained in coupled NLS systems.  Since our formula is given by an
algebraic form, it is possible to investigate dynamics of any order RW solution (for instance the explicit figure for 20-th or higher order RW) by computer soft. These results will further enrich our realization on RW dynamics in many different nonlinear physical systems.
The method provided in this paper can be extended to derive the high-order RW solutions for coupled Hirota equations and even $N$-component NLS system or $N$-wave system with a general case.

Note added. Very recently, Chen and Mihalache presented fundamental and second-order RW solutions in the same coupled model \cite{Chen2}.  Some of their results are overlapped with ours.

\section*{Acknowledgments}
This work is supported by National Natural Science Foundation of
China (Contact No. 11401221, 11405129 ) and Fundamental Research Funds for the Central Universities
(Contact No. 2014ZB0034).

\end{document}